\documentstyle[12pt]{article}

\newcommand{\beq}{\begin{equation}}
\newcommand{\eeq}{\end{equation}}

\begin{document}


\begin{center}
\vskip 0.75cm

{\Huge {\bf Fusion of strings as a hadronic accelerator}}\\
\vskip 0.8cm

N. Armesto, M. A. Braun$^{*}$, E. G. Ferreiro, C. Pajares and Yu. M. 
Shabelski$^{**}$ \\ 
{\sl Departamento de 
F\'{\i}sica de  Part\'{\i}culas,
Universidade de Santiago de
Compostela,\\ 15706--Santiago de Compostela, Spain}

\vskip 1.0truecm
{\Large {\bf Abstract}}
\end{center}
\begin{quotation}

It is shown that the fusion of strings is a source of particle production in
 nucleus--nucleus collisions outside the kinematical limits of nucleon--nucleon
collisions. 
The spectrum of different particles is compared with the high energy data on 
p--A collisions obtaining a reasonable agreement. Results for A--B collisions
at $\sqrt{s}=19.4$ AGeV and $\sqrt{s}=200$ AGeV are given. It is shown that the
fusion of strings can accelerate particles up to the highest energy detected
in cosmic rays without help of any additional cosmic accelerator. Also the
rise of the average shower depth of maximun for cosmic rays in the energy range
between $10^{16}$ eV and $10^{19}$ eV can be explained by the same mechanism
without requiring any change in the chemical composition.

\end{quotation}
\vspace{2cm}

\noindent $^{*}$ 
Visiting Professor IBERDROLA. Permanent address: Department of High Energy
Physics, University of St. Petersburg, 198904 St. Petersburg, 
Russia.\\ 
\noindent $^{**}$ Permanent address: Petersburg Nuclear Physics Institute, 
Gatchina, 188350 St. Petersburg, Russia. 

\newpage

\section{Introduction}
 
Several models of hadronic interactions 
have been very successful in describing
particle production in hadron--hadron, hadron--nucleus and nucleus--nucleus
collisions. Monte Carlo versions of these models are in reasonable agreement
with most of the properties of soft multiparticle production. 
In these models  
(\cite{And87,Cap94,Wer93,Sor89})
strings, chains or pomerons are exchanged between the projectile and 
target. 
The number of strings grows with the energy and with the number of 
nucleons 
of the participant nuclei. In the first approximation strings fragment 
into particles and resonances
in an independent way. The only correlation among strings is due to 
energy--momentum conservation.
However, the interaction between strings becomes 
important with their number
growing. Even at SPS energies both a large number and a large density of
strings are expected. For instance, for S--S and Pb--Pb central collisions the
estimated number of strings are 120 and 1300, respectively, and their densities
3.3 $fm^{-2}$ and 9.8 $fm^{-2}$, respectively. At the relativistic heavy ion
collider (RHIC) and large hadron collider (LHC) energies the number and density
of strings are not negligibly small even for hadron--hadron collisions.

The interaction between strings or the interaction of resonances produced in
the fragmentation of strings have been introduced in some of the models 
(\cite{Paj90,And91,Sor92,Moh92,Aic93,Ran94}). In particular, fusion 
of strings has been
incorporated into the Dual Parton Model (DPM) (\cite{Moh92}) and 
the Quark Gluon String Model
 (QGSM) (\cite{Ame93}). Some of the effects of string
 fusion, like  
strangeness and antibaryon
 enhancement (\cite{Arm94}), reduction of long range
 correlations
(\cite{Ele94}) 
and multiplicity suppression, are in reasonable agreement with the 
existing experimental data. Also  
predictions for the Relativistic Heavy Ion Collider (RHIC) and 
Large Hadron Collider 
(LHC) are avalaible. In this paper we explore another effect of 
string fusion, namely particle
production in nucleus--nucleus collisions outside the kinematical 
limits of nucleon--nucleon
collisions (the so--called cumulative effect).

It is shown that at present avalaible 
energies a non negligible
number of baryons and mesons are produced with momenta greater than 
the ones of the
colliding nucleons. Essentially, the particles produced outside 
the kinematical
limit are protons and neutrons but an appreciable number of $\pi$'s and
$\Lambda$'s are also predicted. 
This effect,
together with the reduction of multiplicities, provides a natural 
explanation of some 
features of cosmic ray data, like the rise of the average shower 
depth of maximum $X_{max}$  
(the amount of air penetrated
by the cascade when it reaches maximum size) (\cite{Bir93,Gai93}) 
with increasing energy  
from $10^{17}$ eV to 
$10^{19}$ eV, and the existence of events with energy above $10^{20}$ 
eV 
(\cite{Bir93,Efi90,Hay94}), higher than the expected
cut--off (\cite{Gre66,Ste68,Sig11}) due to the
scattering of cosmic rays with the microwave radiation background. 
Usually the first feature is explained by an enrichment of protons in 
the composition of
primary cosmic rays (\cite{Gai93}) as energy increases. However, 
as we shall show, if  
the composition of the primary cosmic rays is 
kept fixed in the energy range between 
$10^{17}$ eV and
$10^{19}$ eV, string fusion leads to a suppression of the multiplicity 
similar
to the one produced by changing heavy nuclei (Fe) by protons in the composition 
of the primary.
 On the other hand, since 
the momentum of a fused string is a sum of the momenta
 of its ancestor strings, 
 it is possible to obtain particles with more energy than the
initial nucleon--nucleon energy. As in a considerable part of events the primary
would be iron, several string could fuse in central Fe--Air collisions.  
Therefore,
 the observed cosmic ray events 
with energy above
$10^{20}$ eV may actually correspond to three or four times less initial 
energy per nucleon  
than the one apparently measured.
 Reducing the energy by a factor three, the attenuation mean free path can grow
more than an order of magnitude.
 String fusion could make these events compatible with 
the existence of the above mentioned cut--off. 

\section{String fusion in the Monte Carlo code}

To study the particle production outside the kinematical limits of 
nucleon--nucleon collisions we use a Monte Carlo code based on the QGSM, 
in which the fusion of strings has
been incorporated (\cite{Ame93}). A detailed description of the Monte 
Carlo String Fusion Model (SFMC) and
comparison with experimental data can be found in Refs. 
\cite{Ame93,Arm94,Ele94}. 
 A hadron or nucleus collision is assumed to be an interaction between 
clouds of partons formed long
before the collision. Without string fusion partons are assumed to 
interact only once. Each
parton--parton interaction leads to the creation of colour strings. 
Since both the projectile and the
target must remain colourless, strings have to be formed in 
pairs. For instance, in nucleon--nucleon collisions at moderate energies a
pair of strings is formed between a valence diquark of the projectile and a
valence quark of the target and between a valence quark of the projectile and a
valence diquark of the target. As the energy increases, pairs of strings are
also formed between sea quarks (antiquarks) of the projectile and sea 
antiquarks (quarks) of the target.
 Hadrons and nuclei are
considered on the same footing. The nuclear wave function is taken as 
a convolution of the parton
distribution in a nucleon with the distribution of nucleons in the 
nucleus. For the distributions of partons and nucleons we take
 gaussian (centered in each nucleon) and 
 Wood--Saxon
shapes respectively.

 Strings fuse when their transverse positions come within
a certain interaction area, which is fixed previously to describe 
correctly the strangeness enhancement (\cite{Arm94}) shown by the data on
nucleus--nucleus collisions.
Fusion can take place only when the rapidity intervals of the strings
overlap.  It is formally described by allowing partons to interact several
times, the number of interactions being the same for projectile and 
target. The
quantum numbers of the fused string are determined by those of the interacting 
partons and
its energy--momentum is the sum of the energy--momentum of the ancestor 
strings.
The
colour charges of the fusing string ends sum into the  colour charge  
of the
resulting string ends according to the $SU(3)$ composition laws. 
In particular,
two triplet strings fuse into an antitriplet  and a sextet string, with
probabilities 1/3 and 2/3 respectively. A triplet and an antitriplet
string give rise to a singlet state and an octet string with 
probabilities 1/9
and 8/9 respectively. In present calculations only fusion of two strings
 is taken into account.

A quasi--classical picture of the decay of colour strings is assumed in which
pairs of oppositely colour partons are produced in the string colour field,
which neutralizes this field and leads to string breaking. The new sextet and
octet strings are supposed to break with the production of two (anti)
quark complexes with the same colour charges $Q$ and $-Q$
 as those of the ends of
the string. 
The created (anti) quarks have arbitrary flavours and masses chosen as the
corresponding constituent masses. The probability rate for the constant colour
field of two opposite charges $Q$--$\overline{Q}$ to create a parton pair with
the same colour charges $Q$--$\overline{Q}$ and transverse mass $M_{t}$ for
unit string length and time is taken by the Schwinger expression 
\beq
W \sim\\ K^2_{[N]}\\ exp(-\pi\\ M^2_t/K_{[N]})
\label{eq1}\ \ ,
\eeq
where $K_{[N]}$ is the string tension for the
$[N]$
$SU(3)$ representation proportional to the corresponding quadratic Casimir
operator $C^2_{[N]}$. In our case
\beq
C^2_{[3]}=4/3,\ \ C^2_{[6]}=10/3,\ \ C^2_{[8]}=3\ \ .
\eeq
The fragmentation of the fused strings produces more baryons and antibaryons,
especially strange ones, than in the case of the fragmentation of the original
strings. These enhancements are accompanied by a strong reduction of particle
production in the central rapidity region.

\section{Cumulative particle production}

Particle 
production outside the nucleon--nucleon kinematical limits is 
a well known effect, called
 cumulative effect, studied both theoretically and experimentally 
(\cite{Bal74,Str81,Efr88,Bra94,Bay79}). However at high enough 
energies, where the string picture can be applied, there are only data from 
one
collaboration at 400 GeV/c (\cite{Bay79}), with incoming protons against 
nuclei: Li, Be, C, 
Al, Cu and Ta. 
We have generated 10000 events for each p--A collision in our Monte Carlo code.
The results together with the experimental data for 
the invariant differential cross sections for production
of protons, positive pions and positive kaons are shown
in Tables 1 and 2.

If the
invariant differential cross section is parametrized as a function of $x_F$ 
in the exponential
form:
\beq
\sigma\sim exp(-bx)
\eeq
the slope $b$ 
is in the range 5--6 for all cases in agreement with the experimental data.
 If the fusion of strings is not included the obtained $b$ value is in the 
range 10--11 quite far from the data.

The fusion of strings describes also rightly the dependence of the inclusive
cross section with the kinetic energy. As an example, in Fig. 1, it is shown
the experimental data for the positive pion inclusive cross section for p--Ta
collisions together the results with and without fusion of strings.

The A--dependence of the cumulative effect is usually parametrized as 
$A^{\alpha(x)}$. The value of $\alpha(x)$ has been proposed (\cite{Efr88}) 
to be the function
\beq
\sigma\sim A^{\alpha(x)}, \ \alpha(x)=1+1/3\ (x-1).
\eeq
which is plotted in Fig. 2 together the $\alpha(x)$ obtained from the
experimental data which is very close to the above straigth line. Also the 
$\alpha(x)$ obtained from the fusion of strings is close to the experimental
data, indeed it fits into the line $\alpha(x)=1+0.4
(x-1)$.

From all these comparisons we can conclude that a reasonable agreement for 
p--A cumulative particle spectrum is obtained. Notice that we do not have any
free parameter. We could improve the comparison by introducing minor effects
like rescattering or a more detailed nucleon and quark correlations (other 
than Fermi motion, which has already been included in our code). However
our goal is not to obtain a perfect fit but just to check if the string fusion
works reasonable well.

Once it has been checked that our model is consistent with the hadron--nucleus
experimental data, we turn to nucleus--nucleus collisions
, simulating 10000  
central S--S collisions and 1000 central Pb--Pb collisions 
 at
 $\sqrt{s}=19.4$ AGeV.
 Also central Pb--Pb collisions at RHIC energies 
($\sqrt{s}=200$ AGeV) have been simulated. 
Distributions of baryons and
mesons in central S--S collisions at SPS energies with $x_F$ 
larger 
than 1 are  shown in Fig. 3.
 In Fig. 4 the separated spectra of protons, neutrons, lambdas, 
$K^0$, $K^+$, $K^-$,
 $\pi^0$, $\pi^+$ and $\pi^-$ are presented.
Fig. 5 and Fig. 6 show the same distributions for the case of central Pb--Pb
collisions at $\sqrt{s}=19.4$ AGeV.  
 The results for Pb--Pb collisions at $\sqrt{s}=200$ AGeV are very
 similar to the ones at $\sqrt{s}=19.4$ AGeV. In 1000 events  2015 
particles are found with  
$|x_F|$ larger than 1 to compare with 1783 at $\sqrt{s}=19.4$ AGeV. 
 This small change is due to a moderate increase of the number of 
strings with energy. In general, it is seen that many baryons and mesons are
produced with $x_F$
larger
than 1.

\section{String fusion at ultrahigh cosmic ray energies}

String fusion produces a 
strong supression of multiplicities. In the limit of a
very strong fusion, the multiplicity in hadron-nucleus collisions turns out
independent of $A$ instead of $\sim$ $A^{1/3}$. Also the squared dispersion 
$D^{2}$ behaves like $A^{-1/3}$ instead of $A^{1/3}$. At finite energies the
 fusion of strings is not strong and it is not expected such a behaviour,
however the reduction of multiplicities is quite
sizable for heavy ion collisions at RHIC and LHC energies. For instance,
simulations done for ALICE detector for LHC
indicate that the multiplicity is a factor between 2.5 and 4 less than
the one obtained in the models without fusion.

This suppression of 
multiplicities can
explain the rise of the average shower depth of maximum in cosmic rays 
as the energy increases, 
without requiring any change in the chemical composition.
It is usually accepted that there is a change in the cosmic ray chemical
composition between $10^{16}$ eV and $10^{19}$ eV. 
It seems that the composition becomes significantly lighter
with increasing energy, going from a heavy composition at $10^{16}$ eV to
a light one at energies higher than $10^{19}$ eV. 
 The distribution of the shower depth of maximum
as a function of energy has been studied using a simple model 
of two components
(\cite{Gai93}), observing that
 the composition of the primary changes from approximatly  
 75 \% of iron component
and a 25 \% of proton
component at $10^{16}$ eV
 to  
 50 \% of iron
and a 50 \% of proton
at $10^{19}$ eV.
To study this point, we have computed the multiplicities of 
p--Air and Fe--Air interactions with and without string fusion in the
 whole range of energies studied (from $10^{16}$ to $10^{19}$ eV).
 As it can be seen in Fig. 7, with string fusion the 
multiplicity for a constant composition 
of 
 10 \% of proton and 90 \% of iron in the whole range of energy, 
essentially reproduces the multiplicity obtained 
without string fusion for a uniform 
change in the composition from 75 \% Fe and 25 \% proton at $10^{16}$ eV to 
50 \% Fe and 50 \% proton at $10^{19}$ eV. Thus the string fusion does the
same job as the composition change.

Therefore, the change in the energy behaviour of the average 
shower depth of maximum $X_{max}$ can be ascribed to a change in the 
interaction mechanism with
the increasing role of collective effects like string fusion, and not to a 
change in the chemical
composition of the primary cosmic rays. Further studies of this 
point would require combining the code used in this paper with the 
standard codes which describe the full cascade. 
 Work in this direction is in progress.

Once we know that it is reasonable to assume that most of the primary cosmic
rays are iron nuclei, we would like to know how many particles are going to be
accelerated due to the string fusion mechanism and get energy-momentum larger
than the permitted limits in nucleon--nucleon collisions.

To study the case relevant for cosmic rays we simulated 1000
Fe--Air collisions at  $10^{17}$
eV (we used this energy and not $10^{20}$ eV to save computing
time, rendering the simulation reliable). In this sample 198 particles with
$|x_F|$ $>$ 1 were found. The average number of strings was found
to be 225, from
 which 62 joined to form  double strings. As mentioned, our code only
includes fusion of two strings. However we can estimate the number of
strings participating in a triple fusion assuming that the probability
for triple fusion is roughly the square of that for double fusion.
 Then one would expect that 18 strings join to form triple strings
and 4 strings join to form a quadruple string. Other reasonable assumptions
about the probability of triple fusion give similar results. Therefore
the probability of obtaining particles with $|x_F|$ $>$ 2 or even
$|x_F|$ $>$ 3 does not seem to be negligible. Triggering central collisions,
 particles with $|x_F|$ $>$ 4 could even be detected.
The energy around $3 \cdot
 10^{20}$
eV measured in several cosmic ray experiments could then be lowered by
a factor 2 to 4 if the described effect is
present and there are particles in the shower with $|x_F|$ $>$ 2
or $|x_F|$ $>$ 3. This lower
energy
 for the primary may lie below
the energy cut--off due to the scattering of cosmic
rays on the microwave background. Indeed, according to the computation of F. W.
Stecker (\cite{Ste68})
the attenuation mean free path increases more than one order of
magnitude. This fact means that the measured energies are compatible with the
effective cut--off.

\section{Conclusions}

The string fusion mechanism produces
particles with more energy--momentum than the original nucleon--nucleon
collisions. In this way it can be considered as a hadronic accelerator. 
We have shown that string fusion reproduces reasonably well the experimental
data on cumulative effect in proton--nucleus collisions giving a sizable number
of protons, neutrons, lambdas and pions in central nucleus--nucleus collisions
already at $\sqrt{s}=19.4$ AGeV. The experimental confirmation of this at SPS
will be welcome. 

Concerning cosmic rays, string fusion can explain the reduction of
multiplicities in the energy range between $10^{16}$ eV and $10^{19}$ eV
without requiring a change in the chemical composition. Also, string fusion
provides a natural hadronic accelerator to reach energies above $10^{20}$ eV
without requiring any additional unusual cosmic accelerator.
 Our predictions can be checked  in future heavy ion experiments
 at the accelerators RHIC, LHC and also in 
 cosmic ray experiments (concretely the Auger 
proyect (\cite{Aug95})).

In conclusion we would like to thank N. S. Amelin, A. Capella, J. W. Cronin, 
G. Parente, J. Ranft and 
E. Zas for useful comments and discussions and the 
Comisi\'on Interministerial de Cienc\'{\i}a y Tecnolog\'{\i}a (CICYT)  
of Spain for financial support. M. A. Braun thanks
IBERDROLA, E. G. Ferreiro thanks
Xunta de Galicia and Yu. M. Shabelski 
the Direcci\'on General de Pol\'{\i}tica Cient\'{\i}fica of Spain 
for finantial support. 
This work was partially supported by the INTAS grant N 93--0079.

\newpage
\noindent{\Large {\bf Table captions}}

\vskip 0.5cm

\noindent{\bf Table 1.} 
Comparison of experimental data (\cite{Bay79}) 
on the invariant differential cross section 
${\sigma} = E \frac{d\sigma}{dp_3}$ 
(GeV mb/(GeV/c)$^3$ sr nucleon)
for $\pi^{+}$ and $K^{+}$ vs $p$  
(GeV/c), laboratory angle $118^{o}$, $P_{lab}$
 = 400 GeV, for p--Li
and p--Ta collisions with  
the String Fusion Model code results, with and without string fusion.

\noindent{\bf Table 2.} 
Comparison of experimental data (\cite{Bay79}) 
on the invariant differential cross section
${\sigma} = E \frac{d\sigma}{dp_3}$ 
(GeV mb/(GeV/c)$^3$ sr nucleon)
for $\pi^{+}$ vs $p$ 
(GeV/c), laboratory angle $118^{o}$, $P_{lab}$
 = 400 GeV, for p--Li
and p--Ta collisions with
the String Fusion Model code results, with and without string fusion.

\newpage
\noindent{\Large {\bf Figure captions}}

\vskip 0.5cm

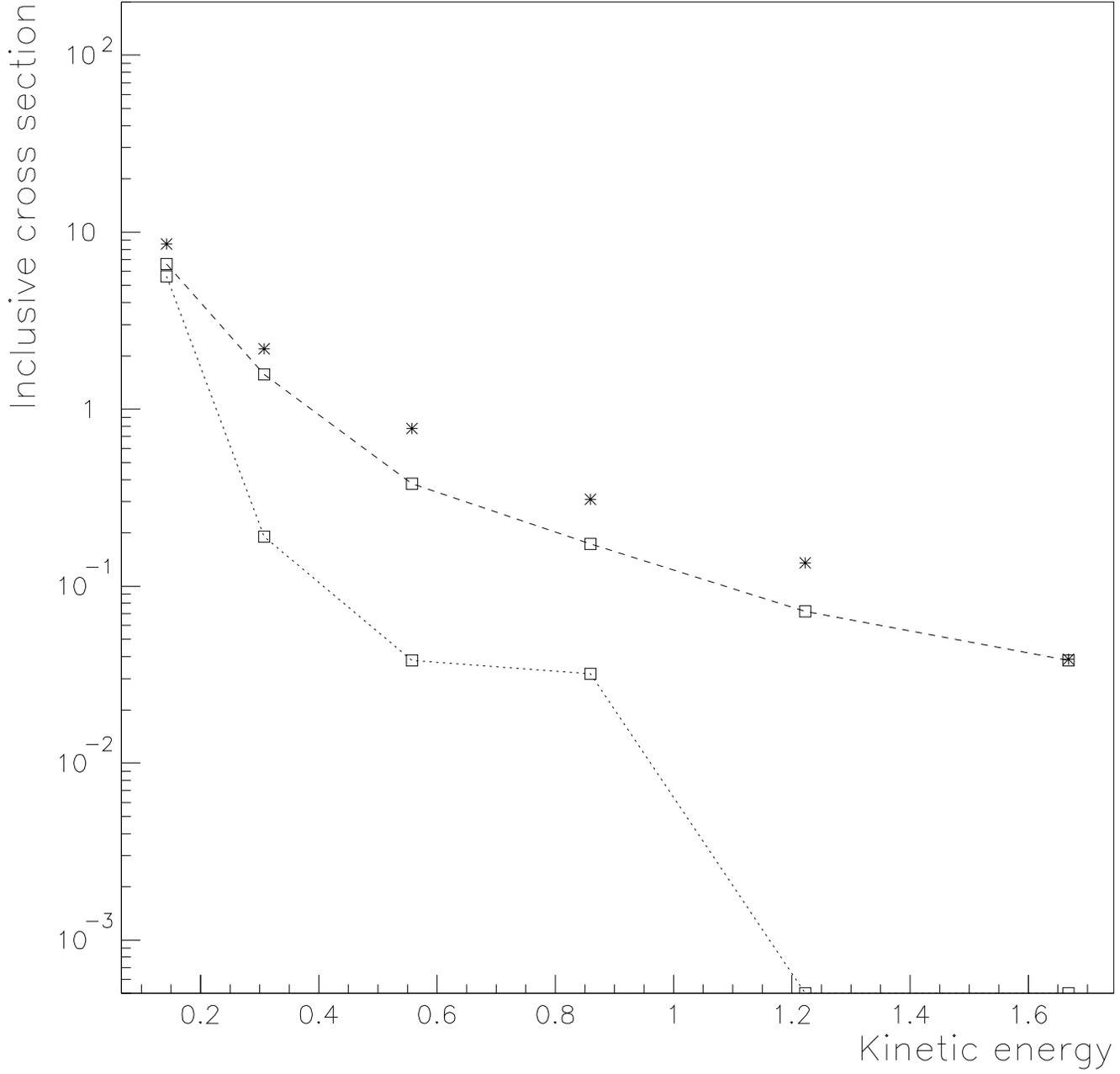
\noindent{\bf Fig. 1.} Comparison of experimental data (\cite{Bay79})
(continuous line)
on the invariant differential cross section
${\sigma} = E \frac{d\sigma}{dp_3}$ for $\pi^{+}$ vs kinetic energy 
(GeV), laboratory angle $118^{o}$, $P_{lab}$
 = 400 GeV, for 
p--Ta collisions with
the String Fusion Model code results, with (upper dashed line) 
and without string fusion. 

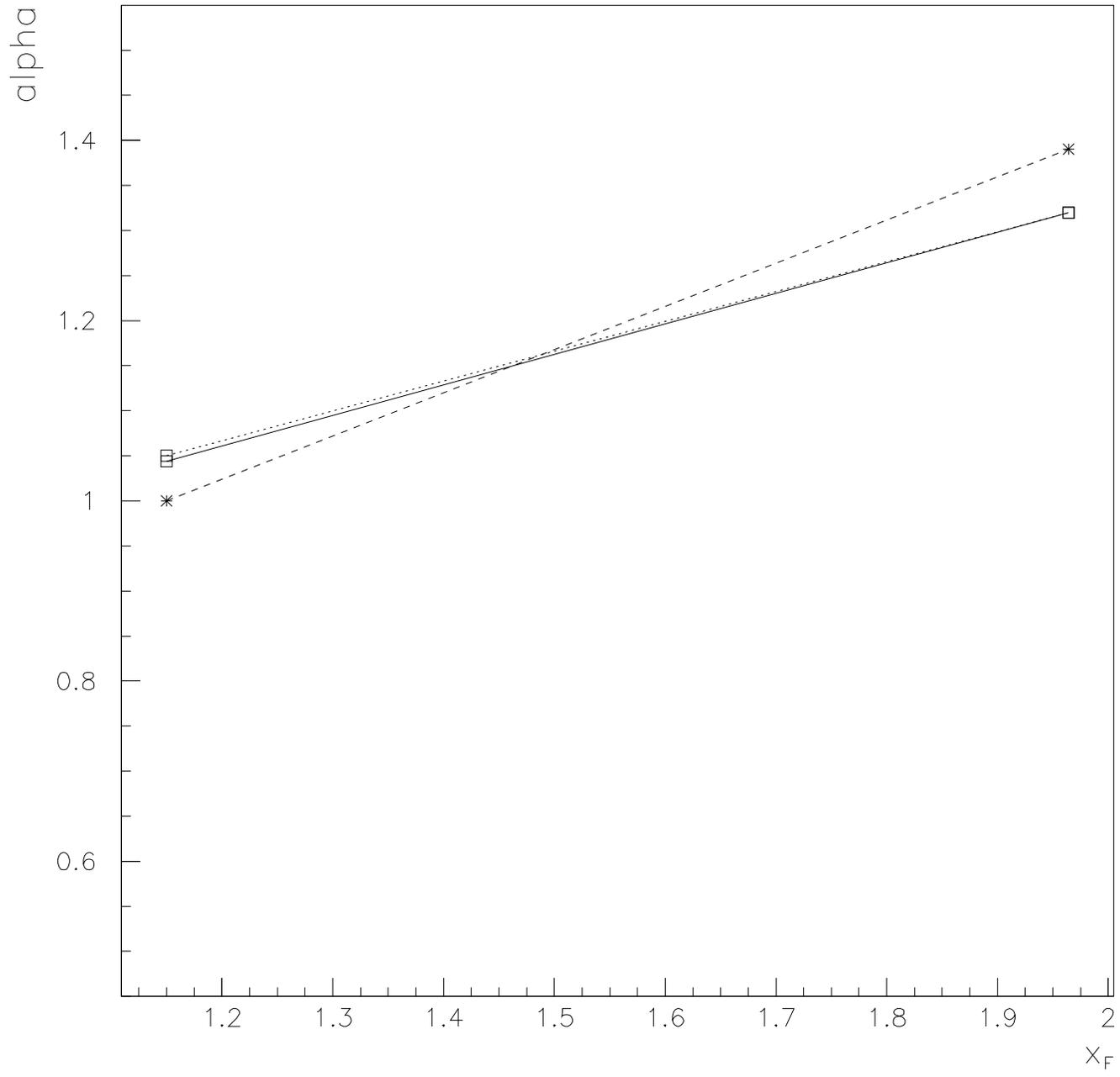
\noindent{\bf Fig. 2.}  ${\alpha}$ vs x. ${\alpha}=1+1/3\ (x-1)$
(continuos line), experimental result (nearer dashed line) and
our results with fusion. 

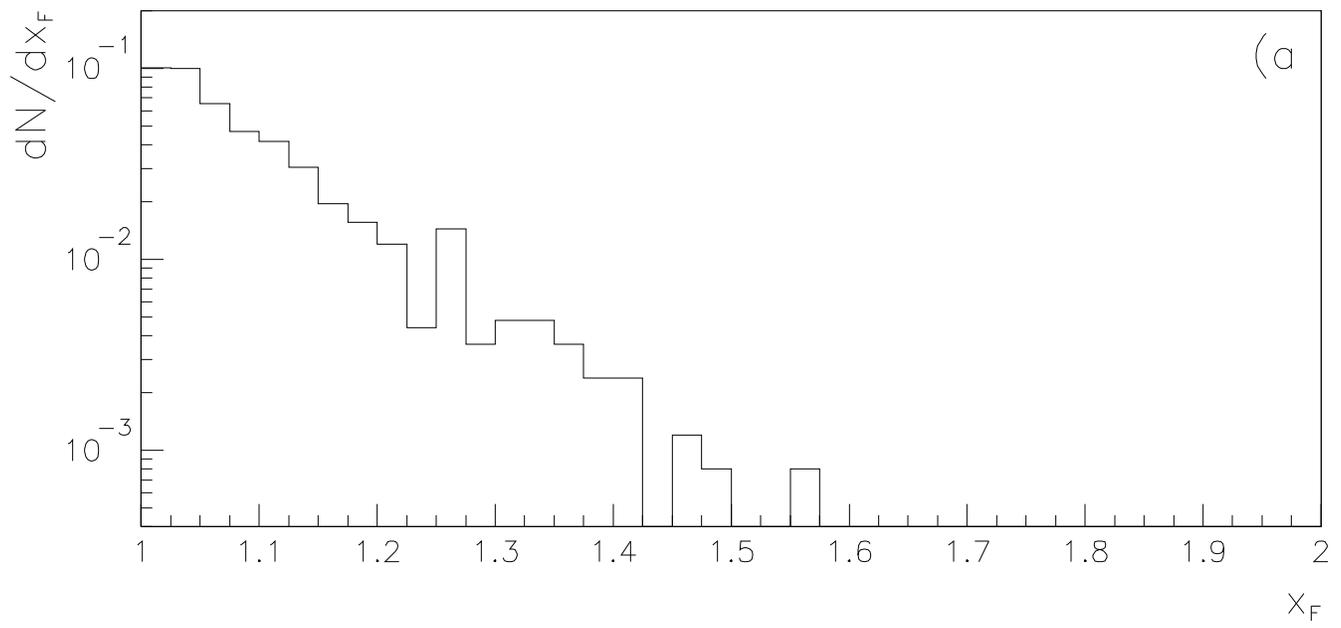
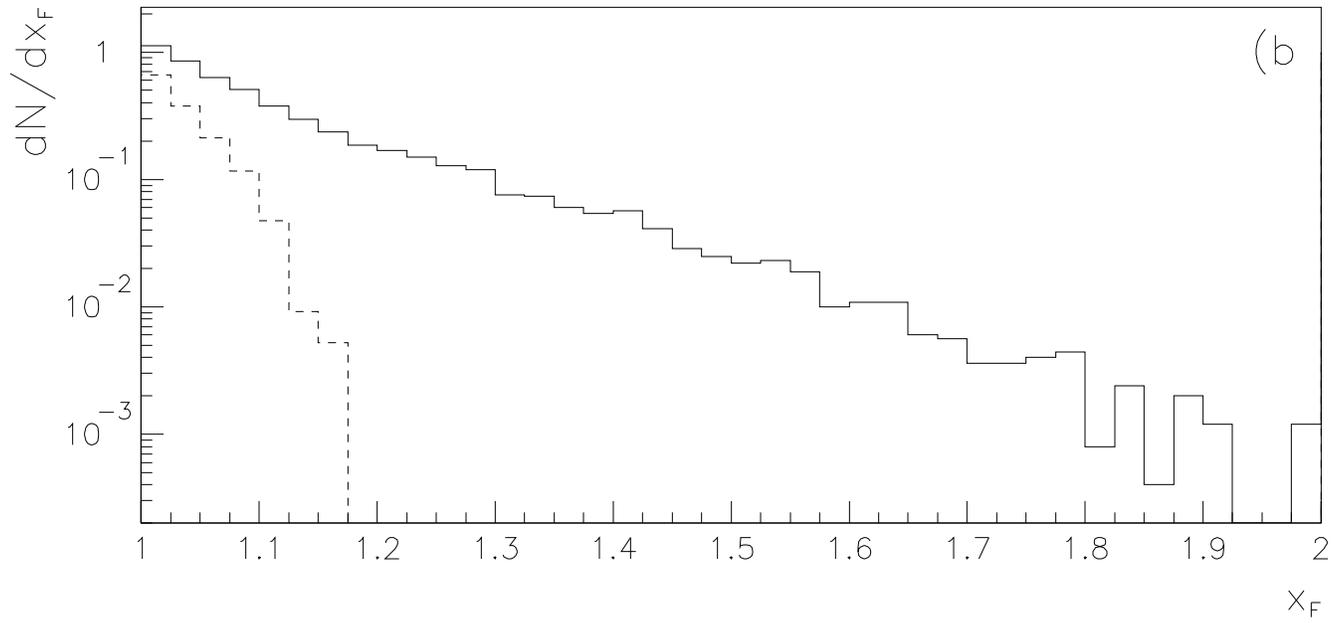
\noindent{\bf Fig. 3.} $x_F$ distributions for $x_F$ $>$ 1 in S--S 
collisions (10000 events) at $\sqrt{s}=19.4$ AGeV of mesons (a) and
baryons (b) with (continuous line) and without (dashed line)
string fusion. No mesons are found in the no fusion case.

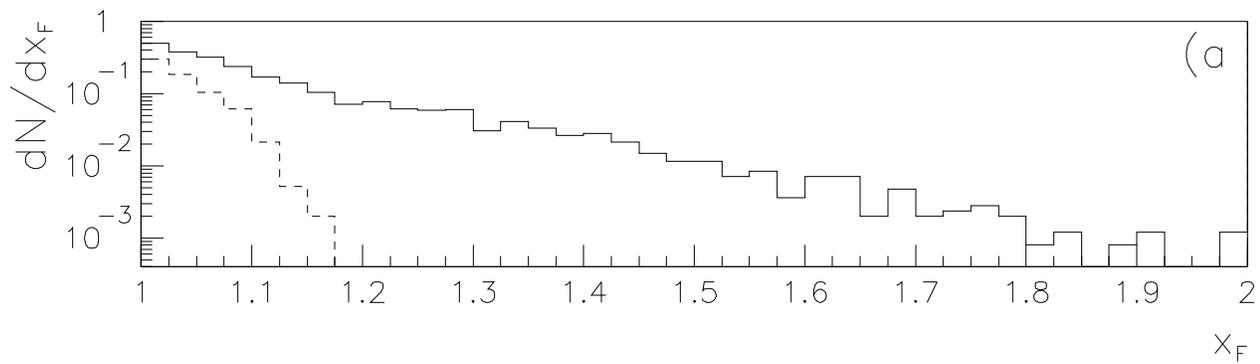
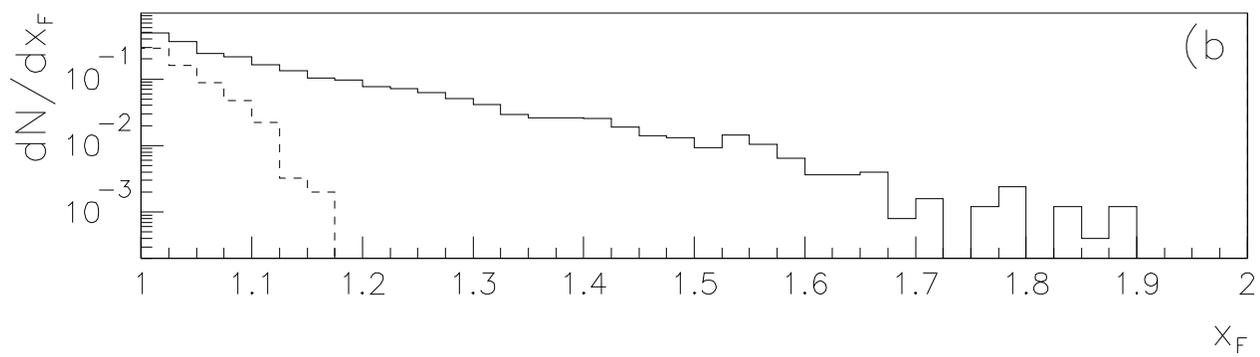
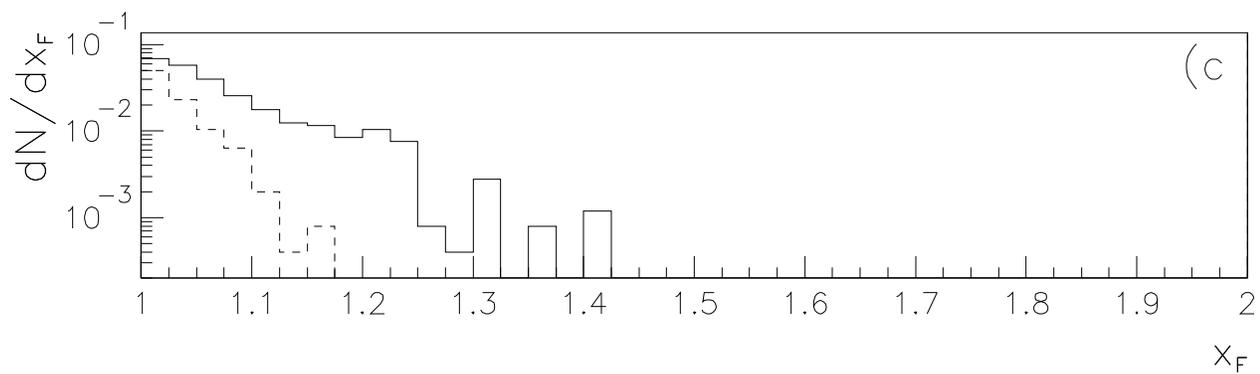
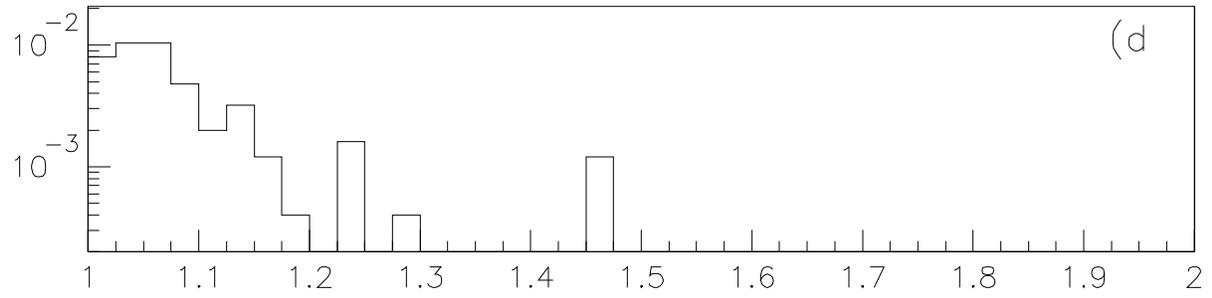
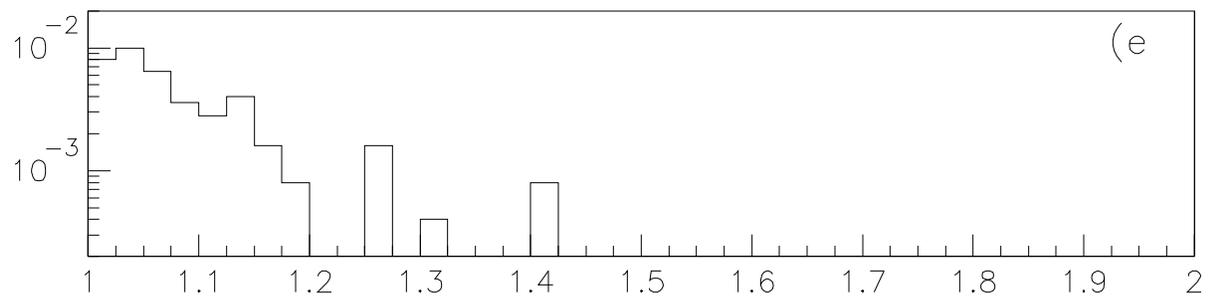
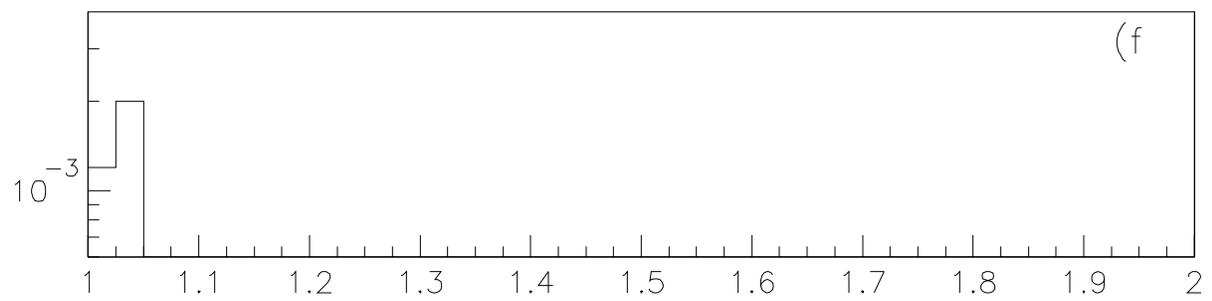
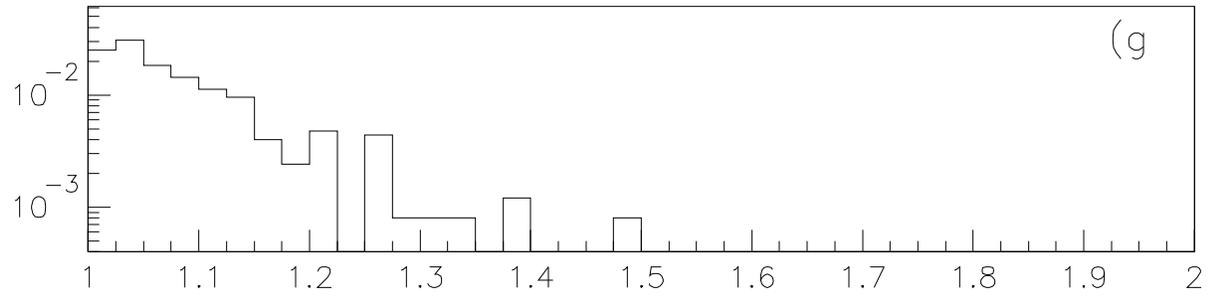
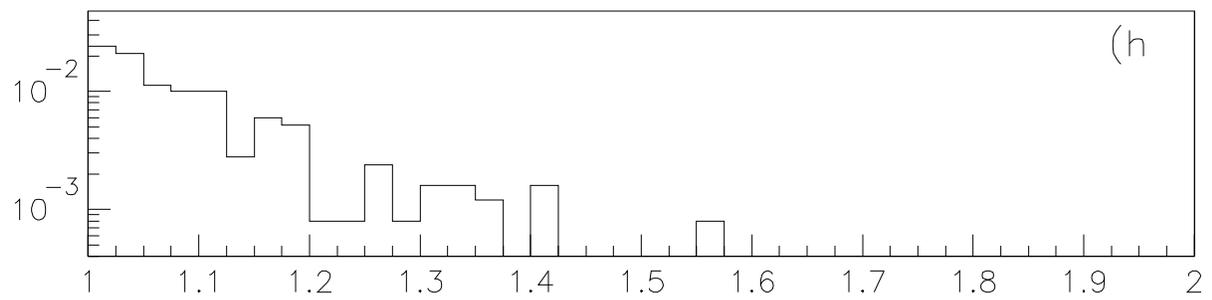
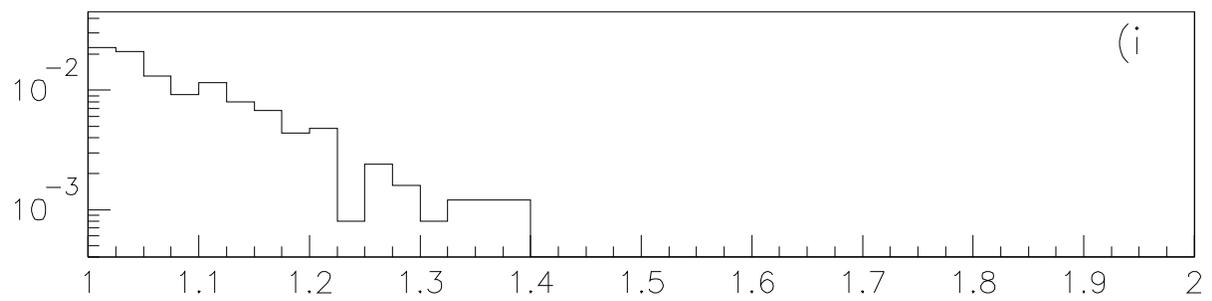
\noindent{\bf Fig. 4.} $x_F$ distributions for $x_F$ $>$ 1 in S--S
collisions (10000 events) at $\sqrt{s}=19.4$ AGeV of 
protons (a), neutrons (b), lambdas (c), $K^0$ (d),
$K^+$ (e), $K^-$ (f), $\pi^0$ (g), $\pi^+$ (h) and $\pi^-$ (i) 
 with (continuous line) and without (dashed line)
string fusion. No mesons are found in the no fusion case.

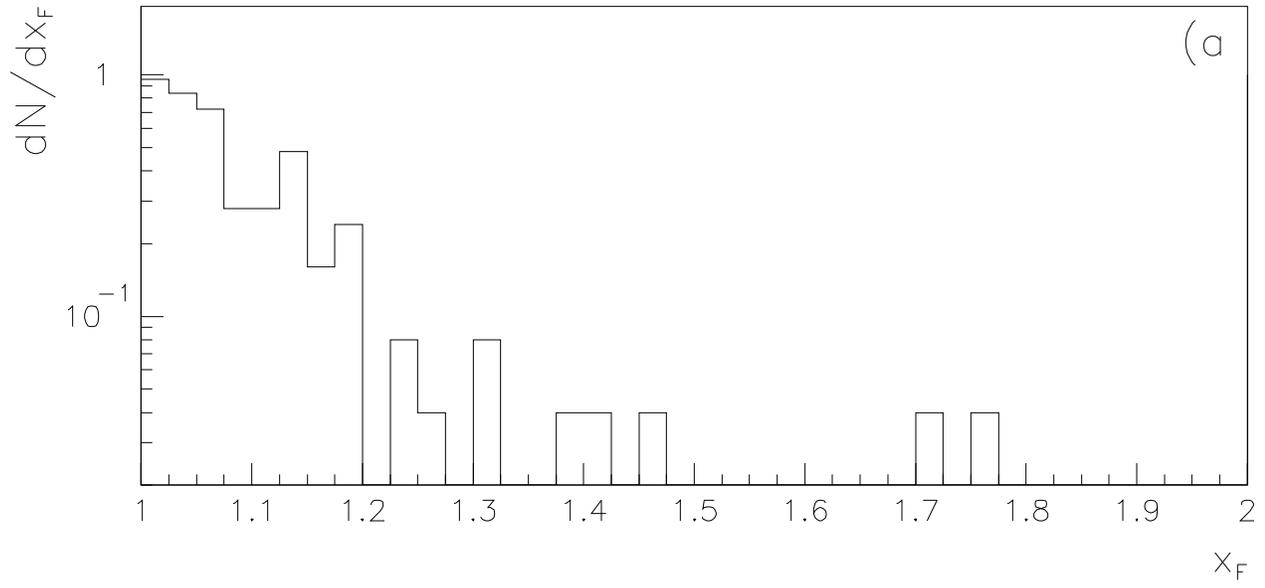
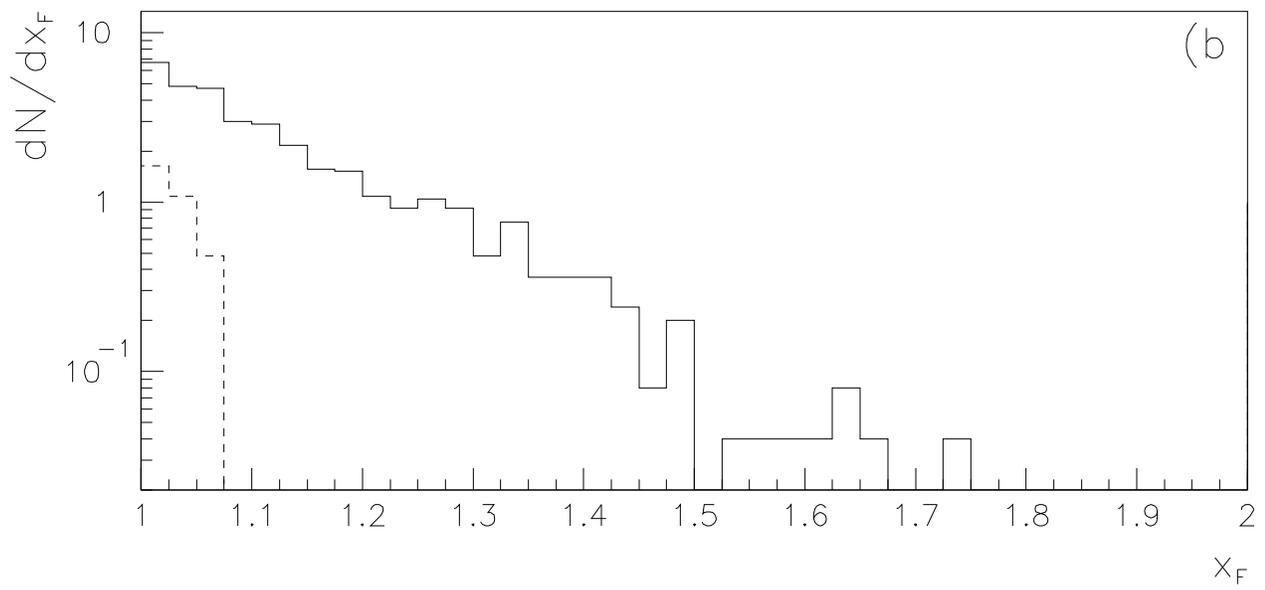
\noindent{\bf Fig. 5.} $x_F$ distributions for $x_F$ $>$ 1 in Pb--Pb 
collisions (1000 events) at $\sqrt{s}=19.4$ AGeV of mesons (a) and
baryons (b) with (continuous line) and without (dashed line)
string fusion. No mesons are found in the no fusion case.

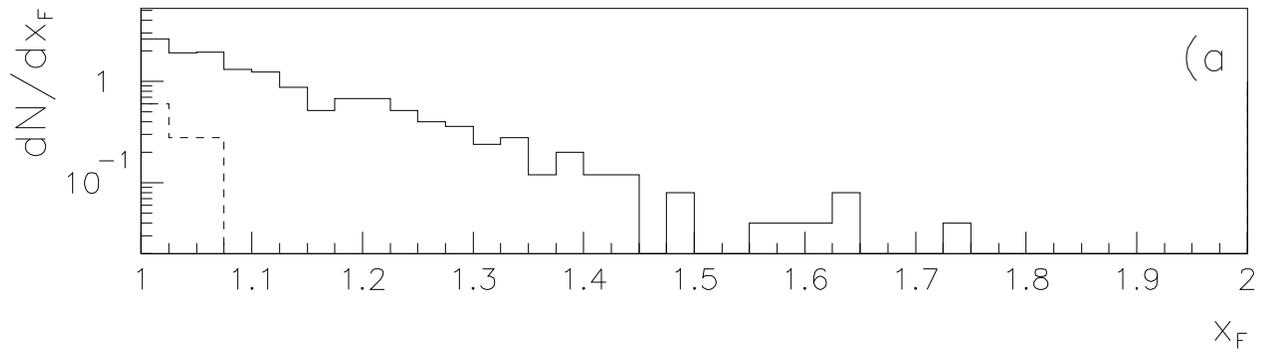
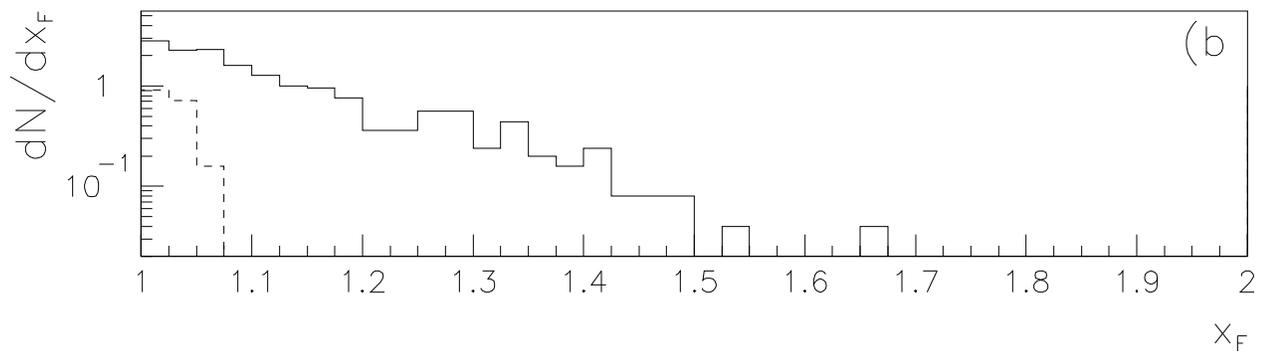
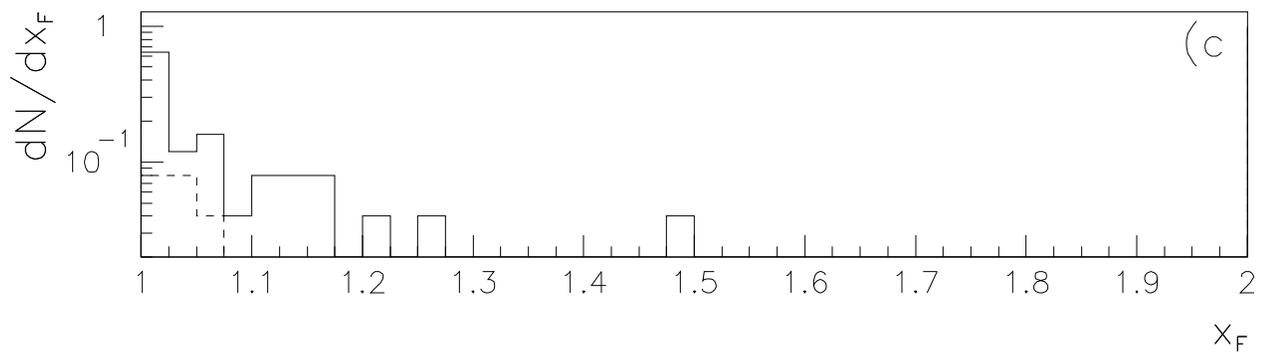
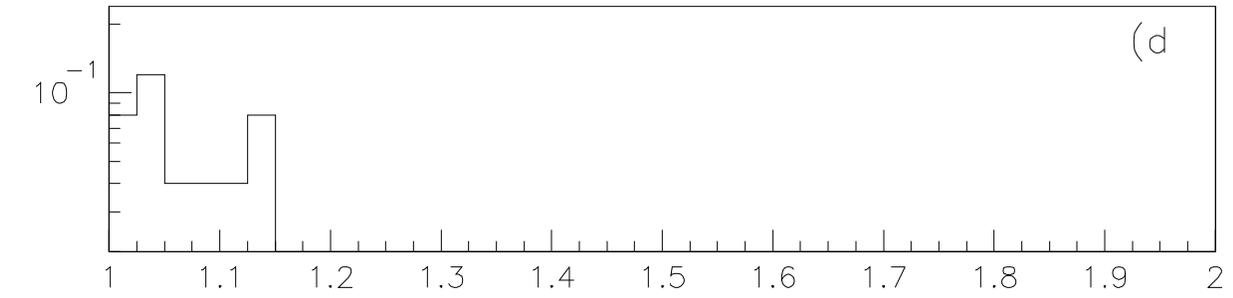
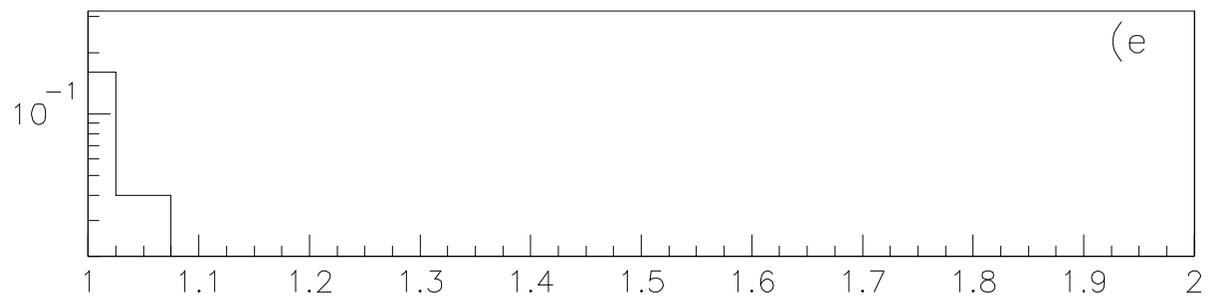
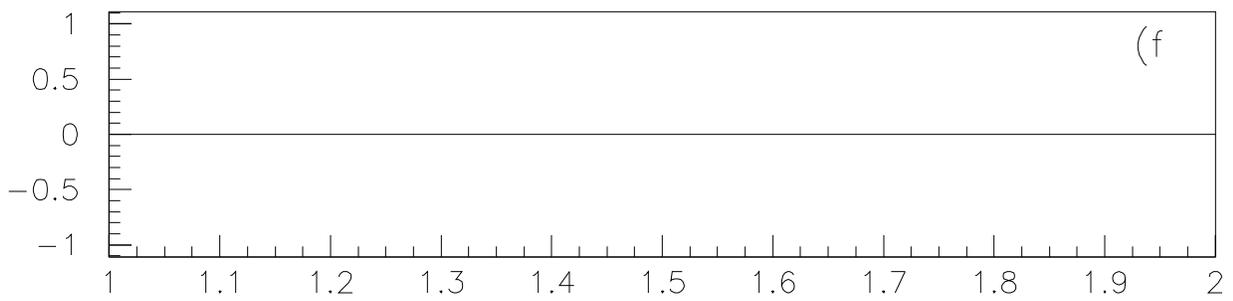
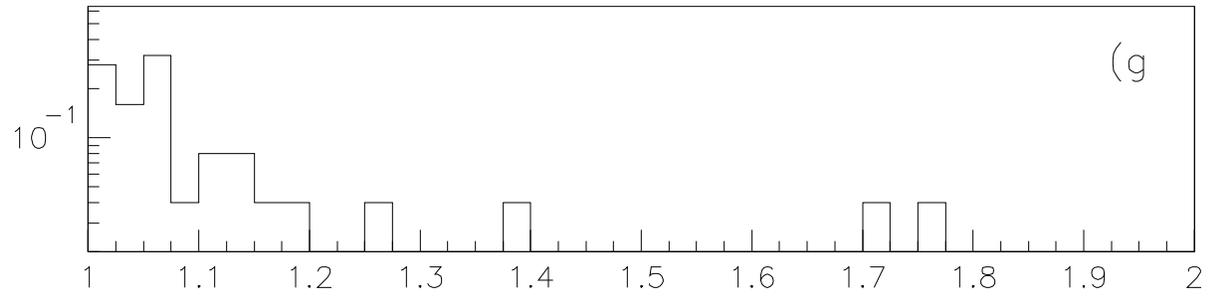
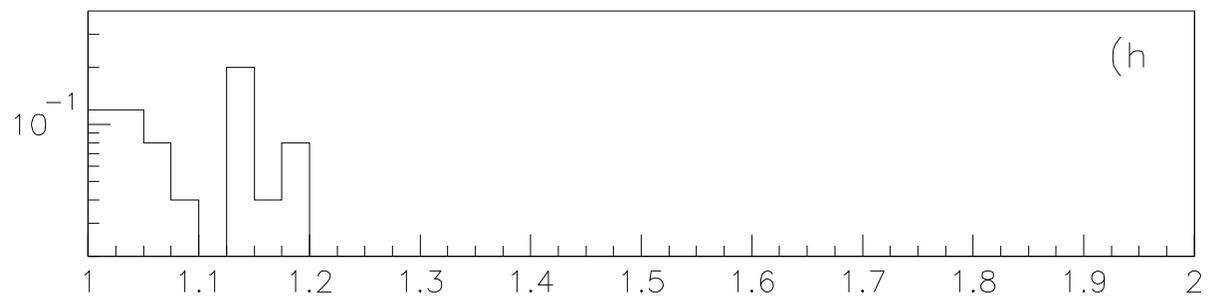
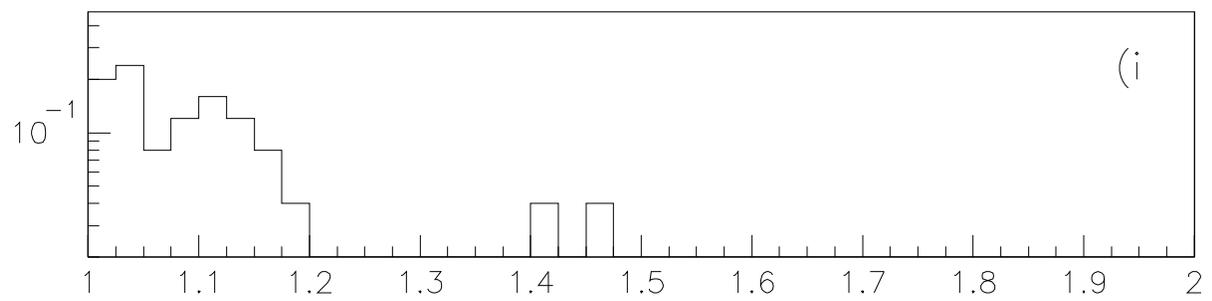
\noindent{\bf Fig. 6.} $x_F$ distributions for $x_F$ $>$ 1 in Pb--Pb 
collisions (1000 events) at $\sqrt{s}=19.4$ AGeV of
protons (a), neutrons (b), lambdas (c), $K^0$ (d),
$K^+$ (e), $K^-$ (f), $\pi^0$ (g), $\pi^+$ (h) and $\pi^-$ (i)
 with (continuous line) and without (dashed line)
string fusion. No mesons are found in the no fusion case.

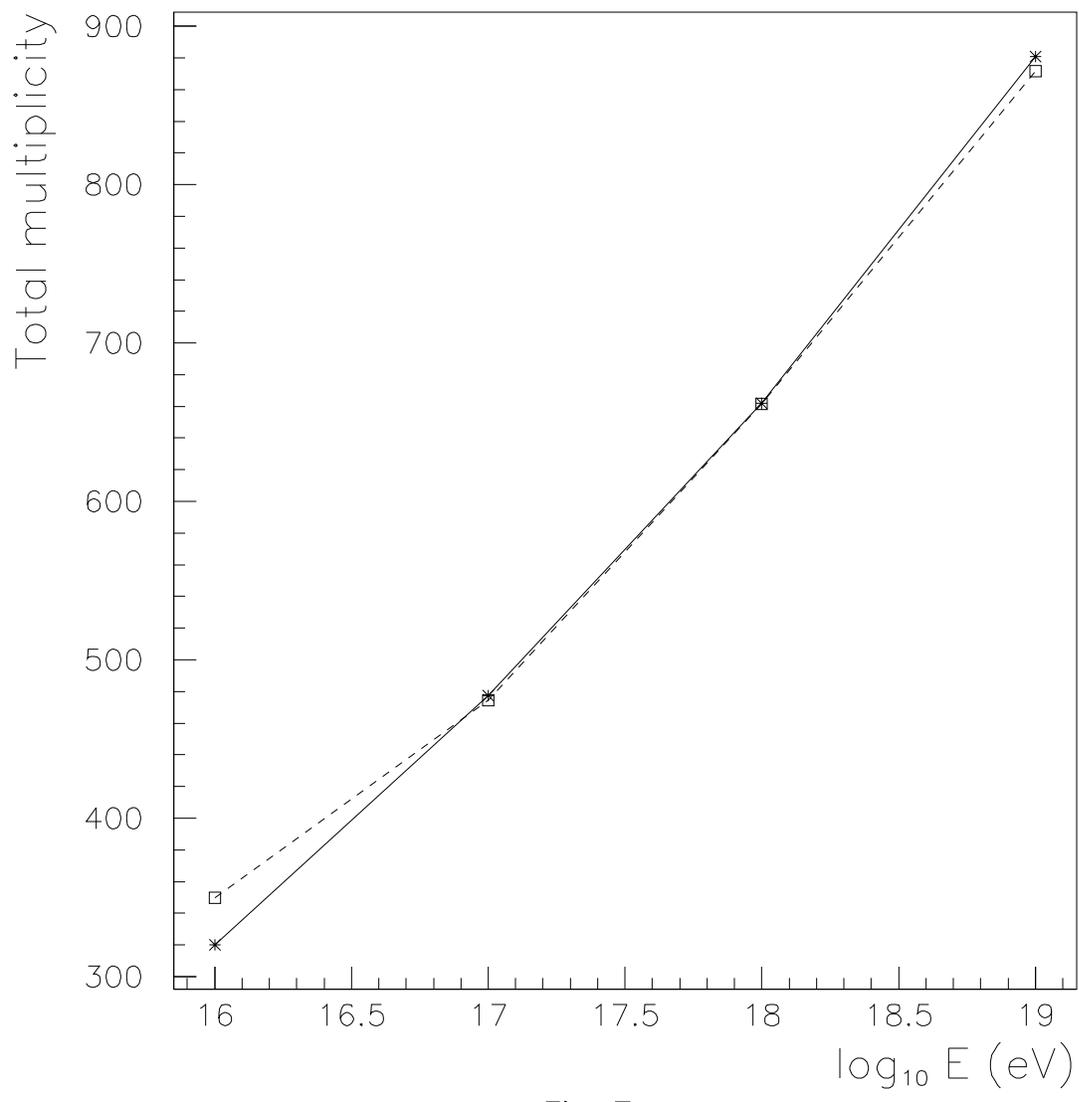
\noindent{\bf Fig. 7.} Total multiplicity dependence on the primary energy
for a fixed composition $<n_t>$ = 0.1 $<n_{p-Air}>$ + 0.9 $<n_{Fe-Air}>$
in the fusion case and a uniform change in the composition from 
$<n_t>$ = 0.25 $<n_{p-Air}>$ + 0.75 $<n_{Fe-Air}>$ at $10^{16}$ eV to
$<n_t>$ = 0.5 $<n_{p-Air}>$ + 0.5 $<n_{Fe-Air}>$ at $10^{19}$ eV
 in the no fusion case (dashed line).

\newpage

\begin{center}
{\bf Table 1}
\vskip 1cm
\begin{tabular}{cccc} \hline
\hline
Reaction & $p-Li$ & $p-Li$ & $p-Li$ \\ \hline
$p$ momentum & Experiment & Without fusion & With fusion  \\ \hline
& ${\sigma}$ for $\pi^{+}$ & ${\sigma}$ for $\pi^{+}$ 
& ${\sigma}$ for $\pi^{+}$ \\ \hline
0.200 & 5.75$\pm$0.79 & 3.53 & 4.77 \\ 
0.293 & 1.89$\pm$0.26 & 0.314 & 1.41 \\ 
0.381 & 0.672$\pm$0.046 & 0.07 & 0.38 \\ 
0.474 & 0.217$\pm$0.016 & 0 & 0.34 \\ 
0.580 & (0.509$\pm$0.044)$10^{-1}$ & 0 & 0.094 \\ 
0.681 & (0.128$\pm$0.012)$10^{-1}$ & 0 & 0.009 \\ \hline
\hline
Reaction & $p-Ta$ & $p-Ta$ & $p-Ta$ \\ \hline
$p$ momentum & Experiment & Without fusion & With fusion  \\ \hline
& ${\sigma}$ for $\pi^{+}$ & ${\sigma}$ for $\pi^{+}$
& ${\sigma}$ for $\pi^{+}$ \\ \hline
0.200 & 8.57$\pm$1.14 & 5.65 & 6.60 \\ 
0.293 & 2.20$\pm$0.31 & 0.19 & 1.57 \\ 
0.394 & 0.78$\pm$0.068 & 0.038 & 0.38 \\ 
0.489 & 0.309$\pm$0.032 & 0.032 & 0.173 \\ 
0.583 & 0.135$\pm$0.017 & 0 & 0.072 \\ 
0.680 & (0.386$\pm$0.076)$10^{-1}$ & 0 & 0.038 \\ \hline
& ${\sigma}$ for $K^{+}$ & ${\sigma}$ for $K^{+}$ 
& ${\sigma}$ for $K^{+}$ \\ \hline
0.539 & (0.241$\pm$0.100)$10^{-1}$ & 0 & 0.037 \\
0.584 & (0.372$\pm$0.763)$10^{-2}$ & 0 & 0 \\ \hline
\end{tabular}
\end{center}

\newpage

\begin{center}
{\bf Table 2}
\vskip 1cm
\begin{tabular}{cccc} \hline
\hline
Reaction & $p-Li$ & $p-Li$ & $p-Li$ \\ \hline
$p$ momentum & Experiment & Without fusion & With fusion  \\ \hline
& ${\sigma}$ for protons & ${\sigma}$ for protons
& ${\sigma}$ for protons \\ \hline
0.385 & 4.17$\pm$0.23 & 1.24 & 2.82 \\ 
0.476 & 1.76$\pm$0.11 & 0 & 1.01 \\ 
0.581 & 0.61$\pm$0.04 & 0 & 0.47 \\ \hline
\hline
Reaction & $p-Ta$ & $p-Ta$ & $p-Ta$ \\ \hline
$p$ momentum & Experiment & Without fusion & With fusion  \\ \hline
& ${\sigma}$ for protons & ${\sigma}$ for protons
& ${\sigma}$ for protons \\ \hline
0.395 & 29.9$\pm$1.5 & 15.1 & 22.3 \\ 
0.490 & 13.2$\pm$0.7 & 0 & 12.57 \\ 
0.585 & 5.2$\pm$0.3 & 0 & 2.5 \\ \hline
\end{tabular}
\end{center}

\end{document}